\documentclass[10pt,conference]{IEEEtran}\IEEEoverridecommandlockouts
\usepackage{epsfig,graphics,subfigure,psfrag,amsmath,cases}
\usepackage{latexsym,amssymb,amsmath,epsfig,subfigure,algorithm}
\usepackage{algorithmic}
\usepackage{color}
\usepackage{url}
\usepackage{scrtime}

\author{\authorblockN{ Derrick Wing Kwan Ng\authorrefmark{1}, Ernest S. Lo\authorrefmark{2}, and Robert Schober\authorrefmark{1}\thanks{This work was supported in part by the AvH Professorship Program of the Alexander von Humboldt Foundation.}}
\authorrefmark{1}Institute for Digital Communications, Universit\"at Erlangen-N\"urnberg, Germany\\
\authorrefmark{2}Centre Tecnol\`{o}gic de Telecomunicacions de Catalunya - Hong Kong (CTTC-HK) \vspace*{-6mm}
}

\title{Energy-Efficient Power Allocation in OFDM Systems with Wireless Information and Power Transfer\vspace*{-5mm} }

\date{\thistime,\,\today}

\newtheorem{Thm}{Theorem}

\newcommand{\abs}[1]{\lvert#1\rvert}

\begin{document}

\maketitle

\begin{abstract}
This paper considers an  orthogonal frequency division multiplexing
 (OFDM) downlink point-to-point system with simultaneous wireless information and power transfer. It is assumed that the receiver is able to harvest energy from noise, interference, and the desired signals.
 We study the design of power allocation algorithms  maximizing
the energy efficiency of data transmission (bit/Joule delivered to
the receiver). In particular, the algorithm design is formulated as a high-dimensional
non-convex optimization problem  which takes into account the circuit
power consumption, the minimum required data rate, and a constraint on the minimum   power delivered to the receiver. Subsequently, by exploiting the properties of
nonlinear fractional programming, the considered non-convex optimization
problem, whose objective function is in fractional form,  is transformed into an equivalent
optimization problem having an objective function in subtractive form, which enables the
derivation of an efficient iterative power allocation algorithm.
In each iteration, the optimal power allocation solution is derived based on  dual decomposition and  a one-dimensional search. Simulation results illustrate that the proposed
iterative power allocation algorithm converges to the optimal solution, and unveil  the trade-off between energy efficiency, system capacity, and wireless power transfer: (1) In the low transmit power regime,
maximizing the system capacity may  maximize the energy
efficiency. (2) Wireless power transfer can enhance the energy efficiency, especially in the interference limited regime.
\end{abstract}

\renewcommand{\baselinestretch}{0.935}
\large\normalsize

\section{Introduction}
\label{sect1}
 Orthogonal frequency
division multiplexing (OFDM) is  a viable air interface for providing ubiquitous communication services and high  spectral efficiency,  due to its ability to combat frequency selective multipath fading and flexibility in resource allocation.  However, power-hungry circuitries  and the limited energy supply in portable devices remain the bottlenecks in prolonging the lifetime of networks and guaranteeing quality of service.  As a
 result, energy-efficient  mobile communication has received considerable interest from both industry
and academia  \cite{JR:Mag_green}-\nocite{CN:static_power,JR:TCOM_harvesting}\cite{JR:TWC_large_antennas}. Specifically, a considerable number of technologies/methods such as energy harvesting and power optimization have been proposed   in the literature
 for maximizing the energy efficiency  (bit-per-Joule) of wireless communication systems. Energy harvesting is particularly  appealing as it is envisioned to  be a perpetual energy source which provides self-sustainability  to systems.

 Traditionally, energy has been harvested from natural renewable energy sources such as solar, wind, and geothermal heat, thereby reducing substantially the reliance on the energy supply from conventional energy sources. On the other hand,  background radio frequency (RF) electromagnetic (EM) waves from ambient transmitters are also an abundant source of energy  for energy harvesting.     Indeed, EM waves can not only serve as a vehicle for carrying  information, but also for carrying energy (/power) simultaneously  \cite{CN:WIPT_fundamental}-\nocite{CN:Shannon_meets_tesla,CN:MIMO_WIPT}\cite{CN:WIP_receiver}. The utilization of this dual characteristic of EM waves leads to  a  paradigm shift for both receivers design and resource allocation algorithm design. In \cite{CN:WIPT_fundamental} and \cite{CN:Shannon_meets_tesla}, the signal input distribution and the power allocation  were used for achieving a trade-off between information and power transfer for different system settings, respectively. However, in \cite{CN:WIPT_fundamental} and \cite{CN:Shannon_meets_tesla} it was assumed that the receiver is able to decode information and extract power from the same received signal which is not yet possible in practice.  As a compromise solution, the concept of a power splitting receiver was introduced in \cite{CN:MIMO_WIPT} and  \cite{CN:WIP_receiver} for facilitating simultaneous  energy harvesting and information decoding.  The authors of  \cite{CN:MIMO_WIPT} and  \cite{CN:WIP_receiver} investigated the rate-energy regions
 for  multiple antenna and single antenna narrowband systems with  power splitting receivers, respectively. Nevertheless,  the possibly high power consumption of both electronic circuitries and  RF transmission was not taken into account in \cite{CN:WIPT_fundamental}-\nocite{CN:Shannon_meets_tesla,CN:MIMO_WIPT}\cite{CN:WIP_receiver} but may play an important role in designing energy efficient communication systems.

In this paper, we address the above issues. To this end, we
formulate the power allocation algorithm design for energy efficient
communication in OFDM systems with concurrent wireless information and power transfer
 as an optimization problem. The resulting high-dimensional non-convex optimization problem is solved by using an iterative algorithm whose components include nonlinear fractional programming, dual decomposition, and a one-dimensional search. Simulation results illustrate an interesting trade-off between energy efficiency, system capacity, and wireless power transfer.

 \begin{figure*}\vspace*{-4mm}
 \centering
\includegraphics[width=5in]{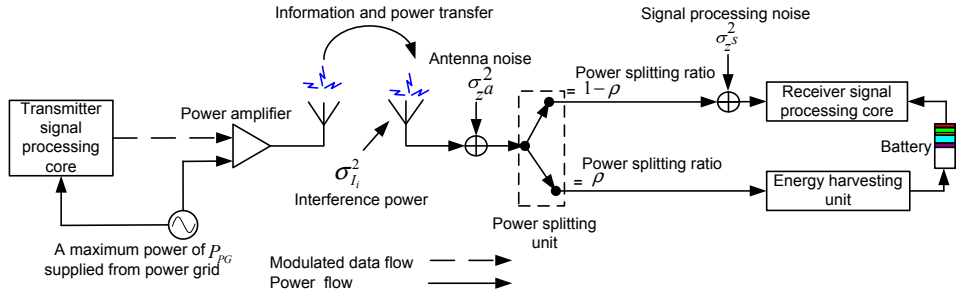}\vspace*{-1mm}
 \caption{OFDM transceiver model for downlink wireless information and power transfer.} \label{fig:system_model}\vspace*{-4mm}
\end{figure*}

\section{System Model}
\label{sect:OFDMA_AF_network_model}
In this section, we present the adopted system model.
\subsection{OFDM Channel Model}
We consider an OFDM system which comprises one transmitter and one receiver. The receiver is able to decode information and harvest energy from noise and radio signals (desired signal and interference signal). All transceivers are equipped with a
single antenna, cf.  Figure \ref{fig:system_model}. The total
 bandwidth of the system is $\cal B$ Hertz and there are $n_F$
subcarriers. Each subcarrier has  a bandwidth $W={\cal B}/n_F$ Hertz. We assume a frequency division duplexing (FDD) system and the
downlink channel gains can be accurately obtained by feedback from the receiver. The channel impulse
response is assumed to be time invariant (slow fading). The downlink
received symbol at the receiver
on subcarrier $i\in\{1,\,\ldots,\,n_F\}$ is
given by
\begin{eqnarray}
Y_{i}=\sqrt{P_{i}l g}H_{i}X_{i}+I_i+Z_{i}^s +Z_{i}^a,
\end{eqnarray}
where $X_{i}$, $P_{i}$, and ${H}_{i}$  are the transmitted
symbol,  transmitted power, and the small-scale fading coefficient  for the link from
the transmitter to the receiver on subcarrier $i$, respectively.
 $l$ and $g$ represent the
path loss and shadowing between the  transmitter and receiver, respectively.
 $Z_{i}^s$ and  $Z_{i}^a$ represent the signal processing and the antenna noises on subcarrier $i$, respectively. $Z_{i}^s$ and $Z_{i}^a$ are modeled as additive white Gaussian noise (AWGN)  with zero mean and variances $\sigma_{z^s}^2$ and $\sigma_{z^a}^2$, respectively,  cf. Figure \ref{fig:system_model}.  $I_i$ is the received co-channel interference signal on subcarrier $i$ with zero mean and variance $\sigma_{I_i}^2$ which is emitted  by an unintended transmitter in the same channel.

\subsection{Hybrid Information and Energy Harvesting Receiver}
\label{sect:receiver}
In practice, the energy harvesting receiver model depends on the specific
implementation. For instance, both electromagnetic induction and electromagnetic radiation  are able to transfer wireless power and information \cite{CN:Shannon_meets_tesla,CN:WIP_receiver}. However, the associated hardware circuitries can vary significantly. Besides, most energy harvesting circuits   suffer from the half-duplex constraint in energy harvesting.   Specifically, the signal used for harvesting energy cannot be used for decoding of the modulated  information \cite{CN:WIP_receiver}. In order to
provide a general model for a receiver which can harvest energy and decode information, we do not assume a
particular type of energy harvesting receiver. Instead, we follow a similar approach as in \cite{CN:WIP_receiver} and
 focus on a receiver which  splits the received signal into two  power streams carrying a proportion of $\rho$ and $1-\rho$ of the total received signal power before any active analog/digial signal processing is performed, cf. Figure \ref{fig:system_model}. Subsequently, the two streams carrying a fraction of $\rho$ and $1-\rho$ of the total received signal power are used for energy  harvesting and decoding the information in the signal, respectively.
   In this paper, we assume a perfect passive power splitter unit which does not consume any power nor introduce any power loss or noise.  Besides, we assume that the receiver is equipped with a  battery with finite capacity for storing the harvested energy. In other words, there is a finite maximum amount of power which can be harvested by the receiver. We note that in practice, the receiver may be  powered
   by more than one energy source and the harvested energy can be used  as a supplement for supporting the energy consumption\footnote{In this paper, we use a normalized  energy unit, i.e., Joule-per-second. Thus,  the terms ``power" and ``energy" are interchangeable in this context.  } of the receiver.

\section{Resource Allocation}\label{sect:forumlation}
In this section, we introduce the adopted system performance metric and formulate the corresponding power allocation problem.
\subsection{Instantaneous Channel Capacity}
\label{subsect:Instaneous_Mutual_information}
In this subsection, we define the adopted system performance
measure. Given perfect channel state information (CSI) at the
receiver, the channel capacity\footnote{Note that the received interference signal $I_i$ on each subcarrier is treated as AWGN which results in a lower bound of the channel capacity and is commonly done in the literature. } between the transmitter  and the receiver on
subcarrier $i$ with channel bandwidth $W$ is given by
\begin{eqnarray}\label{eqn:cap}
C_{i}&=&W\log_2\Big(1+P_i\Gamma_{i}\Big)\,\,\,\,
\mbox{and}\,\,\\
\Gamma_{i}&=&\frac{(1-\rho)l
g\abs{H_i}^2}{(1-\rho)(\sigma_{z^a}^2+\sigma_{I_i}^2)+\sigma_{z^s}^2},
\end{eqnarray}
where $P_i \Gamma_i$ is the received signal-to-interference-plus-noise ratio (SINR) on subcarrier $i$. The \emph{system capacity} is defined as the total
average number of
 bits successfully delivered to the receiver  and is given by
\begin{eqnarray}
 \label{eqn:avg-sys-goodput} && \hspace*{-5mm} U({\cal P}, { \rho})=\sum_{i=1}^{n_F} C_{i},
\end{eqnarray}
where ${\cal P}=\{ P_i \ge 0, \forall i\}$ is the power allocation policy and $\rho$ is the power splitting ratio introduced in Section \ref{sect:receiver}. On the other hand, we take into  account the total power
consumption of the system in the objective
function for designing an energy efficient power allocation algorithm. To this end, we model the power dissipation  in the system as:
\begin{eqnarray}
 \label{eqn:power_consumption}
U_{TP}({\cal P},\rho)\hspace*{-2mm}&=&\hspace*{-2mm}P_C+
 \sum_{i=1}^{n_F}\varepsilon P_i- P_D -P_I \\
 \mbox{where}\label{eqn:Power_harvested_d}\,\,P_D &=& \hspace*{-3mm}\underbrace{\eta\sum_{i=1}^{n_F} P_i l g \abs{H_i}^2\rho}_{\mbox{Power harvested  from desired signal}}\\ \mbox{and}\,\,\label{eqn:Power_harvested_s} P_I &=&\hspace*{-2.65cm}\underbrace{\eta\sum_{i=1}^{n_F}(\sigma_{z^a}^2+\sigma_{I_i}^2)\rho}_{\mbox{Power harvested  from inteference signal and antenna noise}}\hspace*{-0.65cm}.
\end{eqnarray}
$P_C>0$ is the constant  \emph{circuit signal processing power consumption} in both transmitter and receiver which includes the power dissipation in the  digital-to-analog (/analog-to-digital)
converter, digital/analog
filters, mixer, and frequency synthesizer, and is independent of the actual transmitted or harvested power. $\varepsilon\ge 1$ is a constant which accounts for the inefficiency
of the power amplifier. For instance,   $5$ Watts is consumed in the power amplifier  for every 1 Watt of power radiated in the radio frequency (RF) if $\varepsilon=5$; the power
efficiency is $\frac{1}{\varepsilon}=\frac{1}{5}=20\%$. On the other hand, the minus sign in front of $P_D$ in (\ref{eqn:power_consumption}) indicates that a portion of the  power radiated by the transmitter can be harvested by the receiver. $0\le\eta\le1$ is a constant which denotes the efficiency of the energy harvesting unit for converting the radio signal to electrical energy for storage.
Specifically, the term $\eta  l g \abs{H_i}^2\rho$ in (\ref{eqn:Power_harvested_d}) can be interpreted as a \emph{frequency selective power transfer efficiency} for transferring power from the transmitter to receiver on subcarrier $i$. Similarly,  the minus sign in front of $P_I$ in (\ref{eqn:power_consumption}) accounts for the ability of the receiver to harvest power from interference signals and antenna noise. We note that $U_{TP}({\cal P},\rho)>0$ always holds in practical communication systems for the following reasons. First, it can be observed that $\sum_{i=1}^{n_F}\varepsilon P_i\ge \sum_{i=1}^{n_F} P_i> P_D$  due to path loss and the limited energy harvesting efficiency ($\eta\le 1$). Second,  for achieving  a reasonable system performance in communication,  the interference level in the same channel has to be controlled (via regulation) to a reasonable level.  Therefore,  for a typical value of $P_C$,
 $P_C \gg P_I$ is always valid in practice.

The \emph{energy efficiency} of the considered  system is defined as the
total average number of bits/Joule which is given by
\begin{eqnarray}
 \label{eqn:avg-sys-eff} \hspace*{-8mm}U_{eff}({\cal P},\rho)&=&\frac{U_{}({\cal P},\rho)}{U_{TP}({\cal P},\rho)}.
\end{eqnarray}
\subsection{Optimization Problem Formulation}
\label{sect:cross-Layer_formulation}
The optimal power allocation policy, ${\cal P}^*$, ${\rho}^*$,  can be
obtained by solving
\begin{eqnarray}
\label{eqn:cross-layer}&&\hspace*{10mm} \max_{{\cal P}, \rho
}\,\, U_{eff}({\cal P},\rho) \nonumber\\
\notag \hspace*{-5mm}\mbox{s.t.} &&\hspace*{-5mm}\mbox{C1: }
 P_{\max}^{req}\ge P_D +P_I\ge P_{\min}^{req},\notag\\
&&\hspace*{-5mm}\mbox{C2:}\notag\sum_{i=1}^{n_F}P_i\le P_{\max}, \\
&&\hspace*{-5mm}\notag \mbox{C3:}\,\, P_C+\sum_{i=1}^{n_F}\varepsilon P_i\le P_{PG}, \hspace*{3.2mm} \mbox{C4: }\sum_{i=1}^{n_F} C_i\ge R_{\min},\\ &&\hspace*{-5mm}\mbox{C5:}\,\, P_i\ge 0, \,\, \forall i,\hspace*{18.8mm} \mbox{C6:}\,\, 0\le\rho\le 1.
\end{eqnarray}
Variable $P_{\min}^{req}$ in C1 specifies the minimum required
 power transfer to the receiver. $P_{\max}^{req}$  in C1 limits that maximum amount of harvested power because of the finite capacity of the battery. The value of $P_{\max}$ in C2 puts a limit on the transmit spectrum mask to reduce the amount of out-of-cell
interference. C3 is imposed to guarantee that the total power consumption of the system is less than the maximum  power supply from the power grid $P_{PG}$, cf. Figure \ref{fig:system_model}.
 C4   is the
minimum required data rate $R_{\min}$  whose values is provided by the application layer.

\section{Solution of the Optimization Problem} \label{sect:solution}

The first step in solving the non-convex problem in (\ref{eqn:cross-layer})  is to handle the objective function which comprises the ratio of two functions. We note
that there is no standard approach for solving non-convex
optimization problems in general. However, in order to derive an efficient
power allocation algorithm for the considered problem, we
transform the objective function using techniques from nonlinear fractional programming.
\subsection{Transformation of the Objective Function} \label{sect:solution_dual_decomposition}

 For the sake
of notational simplicity, we first define $\mathcal{F}$ as the set of
feasible solutions of the optimization problem in
(\ref{eqn:cross-layer}) and $\{{\cal P},{\cal \rho}\}\in\mathcal{F}$.
  Without loss of generality, we denote $q^*$ as the
maximum energy efficiency of the considered system which is given by
\begin{eqnarray}
q^*=\frac{U({\cal P^*},{\cal \rho^*})}{U_{TP}({\cal
P^*},{\cal \rho^*})}=\max_{{\cal P}, {\cal \rho}}\,\frac{U({\cal P},{\cal \rho})}{U_{TP}({\cal P},{\cal \rho})}.
\end{eqnarray}
We are now ready to introduce the following Theorem which is borrowed from nonlinear fractional programming  \cite{JR:fractional}.
\begin{Thm}\label{Thm:1}
The maximum energy efficiency $q^*$  is achieved if and only if
\begin{eqnarray}\notag
\max_{{\cal P}, {\cal \rho}}&& \hspace*{-2mm}\,U({\cal
P},{\cal \rho})-q^*U_{TP}({\cal P},{\cal \rho})\\
 =&& \hspace*{-2mm}U({\cal
P^*},{\cal \rho^*})-q^*U_{TP}({\cal P^*}, {\cal
 \rho^*})=0,
\end{eqnarray}
\end{Thm}
for $U({\cal P},{\cal \rho})\ge0$ and $U_{TP}({\cal P},{\cal
\rho})>0$.

 \emph{\,Proof:} Please refer to \cite[Appendix A]{JR:TWC_large_antennas} for a  proof similar to the one required for  Theorem 1.

By Theorem \ref{Thm:1}, for any optimization problem
with an objective function in fractional form, there exists an
equivalent optimization problem with an
objective function in subtractive form, e.g. $U({\cal P},{\cal \rho})-q^*U_{TP}({\cal P}, {\cal \rho})$ in the considered
case,  such  that both problem
formulations lead the same optimal power allocation solution.  As a result, we can focus on the equivalent objective function
in the rest of the paper.

\begin{table}[t]\caption{Iterative Power Allocation Algorithm.}\label{table:algorithm}
\vspace*{-5mm}
\begin{algorithm} [H]                    
\caption{Iterative Power Allocation Algorithm }          
\label{alg1}                           
\begin{algorithmic} [1]
\normalsize           
\STATE Initialize the maximum number of iterations $L_{max}$ and the
maximum tolerance $\epsilon$
 \STATE Set maximum energy
efficiency $q=0$ and iteration index $n=0$

\REPEAT [Main Loop]
\STATE Solve the inner loop problem in ($\ref{eqn:inner_loop}$) for
a  given $q$ and obtain power allocation policy $\{{\cal P'}, {\cal \rho'}\}$
\IF {$U({\cal P'}, {\cal \rho'})-q U_{TP}({\cal P'},{\cal
\rho'})<\epsilon$} \STATE  $\mbox{Convergence}=\,$\TRUE \RETURN
$\{{\cal P^*,\rho^*}\}=\{{\cal P',\rho'}\}$ and $q^*=\frac{U({\cal
P'},{\cal \rho'})}{ U_{TP}({\cal P'}, {\rho'})}$
 \ELSE \STATE
Set $q=\frac{U({\cal P'}, {\cal \rho'})}{ U_{TP}({\cal
P'},{\cal \rho'})}$ and $n=n+1$ \STATE  Convergence $=$ \FALSE
 \ENDIF
 \UNTIL{Convergence $=$ \TRUE $\,$or $n=L_{max}$}

\end{algorithmic}
\end{algorithm}
\vspace*{-8mm}
\end{table}

\subsection{Iterative Algorithm for Energy Efficiency Maximization}
In this section, an iterative algorithm (known as the
Dinkelbach method \cite{JR:fractional}) is proposed for solving
(\ref{eqn:cross-layer}) with an equivalent objective function in subtractive form such that the obtained solution satisfies the conditions stated in Theorem 1. The
proposed algorithm is summarized in Table \ref{table:algorithm} and
the convergence to the optimal energy efficiency is guaranteed if  the inner problem (\ref{eqn:inner_loop}) can be solved in each iteration.

\emph{Proof: }Please refer to \cite[Appendix B]{JR:TWC_large_antennas}  for a proof of convergence.

As shown in Table \ref{table:algorithm}, in each iteration in the
main loop, i.e., lines 3--12,  we solve the following optimization problem for a given
parameter $q$:
\begin{eqnarray}\label{eqn:inner_loop}
&&\hspace*{-18mm}\max_{{\cal P},  {\cal \rho}}
\quad\,{U}({\cal P},{\cal \rho})-q{U}_{TP}({\cal P},{\cal \rho}
)\nonumber\\
&&\hspace*{-15mm}\mbox{s.t.} \,\,\mbox{C1, C2, C3, C4, C5, C6}.
\end{eqnarray}

We note that ${U}({\cal P},{\cal \rho})-q{U}_{TP}({\cal P},{\cal \rho}
)\ge 0$ holds  for any value of $q$ generated by Algorithm I. Please refer  to \cite[Proposition 3]{JR:TWC_large_antennas} for a proof.

\subsubsection*{Solution of the Main Loop Problem}
The transformed problem has now an objective function in subtractive form which is less difficult to handle compared to the original formulation. However, there is still an obstacle in tackling the problem.
  The power splitting ratio $\rho$ appears in the capacity equation in each subcarrier which couples
 the power allocation variables and results in a non-convex function, cf. (\ref{eqn:cap}). In order to derive a tractable  power allocation algorithm,
 we have to overcome this problem. To this end, we perform a full search with respect to (w.r.t.) $\rho$. In particular, for a given value of $\rho$,
we optimize the transmit power  for energy efficiency maximization. We repeat the procedure for all possible values\footnote{In practice, we discretize the range of $\rho$, i.e., $[0,1]$, into $M\gg 1$ equally spaced intervals with an interval width of $\frac{1}{M}$ for facilitating the full search.} of $\rho$ and record the corresponding achieved energy efficiencies. At last,  we select that $\rho$ from all the trials which provides the maximum system energy efficiency.  Note that  for a fixed $\rho$, the transformed problem in (\ref{eqn:inner_loop}) is concave w.r.t. the power allocation variables and (\ref{eqn:inner_loop}) satisfies Slater's constraint qualification. As a result, the  search space of the solution set can be reduced from $n_F+1$ dimensions (in problem  (\ref{eqn:cross-layer})) to a one-dimensional search w.r.t. $\rho$ due to the proposed transformation in Theorem \ref{Thm:1} and dual decomposition which will be introduced in the  next section.

Now, we solve the transformed problem for a given value of $\rho$ by exploiting the concavity of the problem.
It can been seen that strong duality holds for the transformed problem for a given value of $\rho$, then solving the dual
problem is equivalent to solving the primal problem \cite{book:convex}.




 \subsection{Dual Problem Formulation}
In this subsection, for a given value of $\rho$, we solve the power allocation
optimization problem by solving its dual.
 For this purpose, we first need
the Lagrangian function of the primal problem. The Lagrangian of  (\ref{eqn:inner_loop}) is given by
\begin{eqnarray}\hspace*{-2mm}&&\notag{\cal
L}( \alpha, \beta,\gamma,\lambda,\theta,{\cal P},{\cal \rho})\\
\hspace*{-5mm}&=&\hspace*{-3mm}\sum_{i=1}^{n_F}
(1+\gamma)C_i\hspace*{-0.5mm}-\hspace*{-0.5mm}q\Big(U_{TP}({\cal P},\rho)\Big)\hspace*{-0.5mm}-\hspace*{-0.5mm}\lambda\Big( P_C+\sum_{i=1}^{n_F}\varepsilon P_i- P_{PG}\hspace*{-1mm}\Big)\notag\\
\hspace*{-2mm}&-&\hspace*{-3mm}\beta\Big(\sum_{i=1}^{n_F}P_i- P_{\max}\Big)-\gamma R_{\min}-\alpha\Big(P_{\min}^{req}-P_D-P_I \Big)\notag\\
\hspace*{-2mm}&+&\hspace*{-3mm}\theta\Big(P_{\max}^{req}-P_D-P_I \Big).
\label{eqn:Lagrangian}
\end{eqnarray}
Here, $\lambda\ge0$ is the Lagrange multiplier connected to C3 accounting for the power usage from
the power grid.
$\beta\ge0$ is the Lagrange multiplier
corresponding to the maximum transmit power limit in C2.  $\alpha\ge 0$  and $\gamma\ge 0$ are the Lagrange multipliers associated with the minimum required power  transfer and the minimum data rate requirement in C1 and C4, respectively. $\theta\ge 0$ is the Lagrange multiplier accounts for the maximum allowed power transfer in C1.   On the other hand, boundary constraints C5
and C6 will be absorbed into the Karush-Kuhn-Tucker (KKT) conditions
when deriving the optimal power allocation solution in the
following.

The dual problem is given by
\begin{eqnarray}
\underset{ \alpha, \beta,\gamma,\lambda,\theta \ge 0}{\min}\ \underset{{\cal
P,\rho}}{\max}\quad{\cal L}( \alpha, \beta,\gamma,\lambda,\theta,{\cal P},{\cal \rho}).\label{eqn:master_problem}
\end{eqnarray}

\subsection{Dual Decomposition Solution }
\label{sect:sub_problem_solution} By Lagrange dual decomposition, the dual
problem can be decomposed into two layers: Layer 1 consists of $n_F$ subproblems with identical structure which can be solved in parallel; Layer 2 is the master problem. The dual problem can be
solved iteratively, where in each iteration the transmitter solves the
 subproblems  by using the KKT conditions for a fixed set of Lagrange multipliers, and the master problem  is solved using the gradient method.

\subsubsection*{Layer 1 (Subproblem Solution)}

 Using standard
optimization techniques and KKT conditions, the optimal power allocation
on subcarrier $i$ for a given $q$  is obtained as
 \begin{eqnarray}\label{eqn:power1}
\hspace*{-3mm}P_{i}^*\hspace*{-2mm}&=&\hspace*{-2mm}\Bigg[\frac{W(1+\gamma)}{\ln(2)\Lambda_i}-\frac{1}{\Gamma_i}\Bigg]^+, \,\forall i,\quad\mbox{where}\\
\hspace*{-3mm}\Lambda_i\hspace*{-2mm}&=&\hspace*{-2mm}q\Big(\varepsilon-\eta\rho l g\abs{H_i}^2\Big)+\lambda\varepsilon\hspace*{-0.5mm}+\hspace*{-0.5mm}\beta\hspace*{-0.5mm}+\hspace*{-0.5mm}(\theta-\alpha)\eta\rho l g\abs{H_i}^2
\end{eqnarray}
and $\big[x\big]^+=\max\{0,x\}$. The power allocation solution in (\ref{eqn:power1}) is in the form of water-filling. Interestingly, the water-level in (\ref{eqn:power1}), i.e., $\frac{W(1+\gamma)}{\ln(2)\Lambda_i}$, is different across different subcarriers due to the \emph{frequency selective power transfer efficiency} described after (\ref{eqn:Power_harvested_d}). On the other hand, Lagrange multipliers $\gamma$ and $\alpha$
force the transmitter to allocate more power for transmission to fulfill the data rate requirement $R_{\min}$ and the minimum power transfer requirement $P_{\min}^{req}$, respectively.

%

\subsubsection*{Layer 2 (Master Problem Solution)}
The dual function is differentiable and, hence, the
gradient method can be used to solve the Layer 2 master problem in
(\ref{eqn:master_problem}) which leads to
\begin{eqnarray}\label{eqn:multipler1}
\hspace*{-5.5mm}\alpha(m+1)\hspace*{-3mm}&=&\hspace*{-3mm}\Big[\alpha(m)-\xi_1(m)\times
\Big( P_D+P_I- P_{\min}^{req}\Big)\Big]^+\hspace*{-1.5mm},\\
\hspace*{-5.5mm}\beta(m+1)\hspace*{-3mm}&=&\hspace*{-3mm}\Big[\beta(m)-\xi_2(m)\times
\Big(P_{\max}-\sum_{i=1}^{n_F} P_i\Big)\Big]^+\hspace*{-1.5mm}, \label{eqn:multipler2}\\
\hspace*{-5.5mm}\gamma(m+1)\hspace*{-3mm}&=&\hspace*{-3mm}\Big[\gamma(m)-\xi_3(m)\times
\Big(\sum_{i=1}^{n_F} C_i -R_{\min}\Big )\Big]^+\hspace*{-1.5mm},\label{eqn:multipler3}\\
\hspace*{-5.5mm}\lambda(m+1)\hspace*{-3mm}&=&\hspace*{-3mm}\Big[\lambda(m)-\xi_4(m)\times
\Big(P_{PG} -P_C-\sum_{i=1}^{n_F}\varepsilon P_i \Big)\Big]^+\hspace*{-2.2mm}, \label{eqn:multipler4}\\
\hspace*{-5.5mm}\theta(m+1)\hspace*{-3mm}&=&\hspace*{-3mm}\Big[\theta(m)-\xi_5(m)\times
\Big(P_{\max}^{req}-P_D-P_I\Big)\Big]^+\hspace*{-1.5mm}, \label{eqn:multipler5}
\end{eqnarray}
where index $m\ge 0$ is the iteration index and $\xi_u(m)$,
$u\in\{1,2,3,4,5\}$, are positive step sizes.  Then, the updated Lagrange multipliers in
(\ref{eqn:multipler1})-(\ref{eqn:multipler5}) are used for solving
the subproblems in (\ref{eqn:master_problem}) via updating the power
allocation solution in (\ref{eqn:power1}).

 Since the transformed  problem  is concave
for given parameters $q$ and $\rho$, it is
guaranteed that the iteration between the Layer 2 master problem and the Layer 1 subproblems converges
to the primal optimal solution of (\ref{eqn:inner_loop}) in the
main loop, if the chosen step
 sizes satisfy the infinite travel condition
 \cite{book:convex,Notes:Sub_gradient}.

After  obtaining the solution of (\ref{eqn:inner_loop}) with the above algorithm for a fixed $\rho$,  we solve (\ref{eqn:inner_loop}) again for another value of $\rho$ until we obtain the  energy efficiency for all considered values of $\rho$.

\section{Results}
\label{sect:result-discussion} In this section, we evaluate the
 performance of the proposed power allocation algorithm using simulations. The TGn path loss model \cite{report:tgn} for indoor communication is adopted with 20 dB directional transmit and receive antennas gains. The distance between the transmitter and receiver is 10 meters. The system bandwidth is ${\cal B}=1$ MHz and the number of subcarriers is $n_F=128$. We assume  a carrier
center frequency of $470$ MHz which will be used by IEEE 802.11 for the next generation of Wi-Fi systems  \cite{report:80211af}. Each subcarrier for RF transmission has a bandwidth of $W=78$ kHz with antenna noise and signal processing noise
powers of $\sigma_{z^a}^2=-128$ dBm and $\sigma_{z^s}^2=-125$ dBm \cite{book:microwave}, respectively.
 The small-scale fading coefficients of
the transmitter and receiver links are generated as independent and identically
distributed (i.i.d.)  Rician random variables with Rician factor equal to 6 dB.
Besides, the received interference at the receiver  on each subcarrier  is generated as i.i.d. Rayleigh random variables with variance specified in each case study.  The shadowing of both the  desired and interference communication links are set to $0$ dB, i.e., $g=1$ for the desired link.
Unless specified otherwise, we assume a static signal processing power
consumption of $P_C=$ 40 dBm, a minimum data rate
requirement of $R_{\min}=10$ Megabits/s, a minimum required power transfer of $P_{\min}^{req}=0$ dBm, a maximum allowed power transfer of $P_{\max}^{req}=20$ dBm, and an energy harvesting efficiency of $\eta=0.8$.  We set $M=1000$ for discretizing the range of $\rho$ into 1000 equally spaced intervals  for performing the full search\footnote{In practice, much smaller values for $M$ (e.g., $M=100$) can be used to reduce complexity at the expense of a small loss in performance. }. On the other hand, we assume a power efficiency of
$38\%$ for the power amplifier used at the transmitter,
i.e., $\varepsilon=\frac{1}{0.38}=2.6316$. The average system
energy efficiency is obtained by counting the number of bits
which are successfully decoded by the receiver over the total power
consumption averaged over multipath  fading.  Note that if the transmitter
is unable to guarantee the minimum required data rate $R_{\min}$ or the minimum required power transfer $P_{\min}^{req}$, we
set the energy efficiency and the system capacity for that channel realization to zero to account for the corresponding failure. For the sake of illustration, we define the interference-to-signal processing noise ratio (INR) as $\frac{ \sigma_{I_{i}}^2}{\sigma_{z^s}^2}$. In the following results, the ``number of iterations'' refers to the
number of outer loop iterations of Algorithm 1 in Table I.
\vspace*{-0.1cm}
\subsection{Convergence of Iterative Algorithm 1 }
\begin{figure}[t]\vspace*{-5mm}
 \centering
\includegraphics[width=3.5 in]{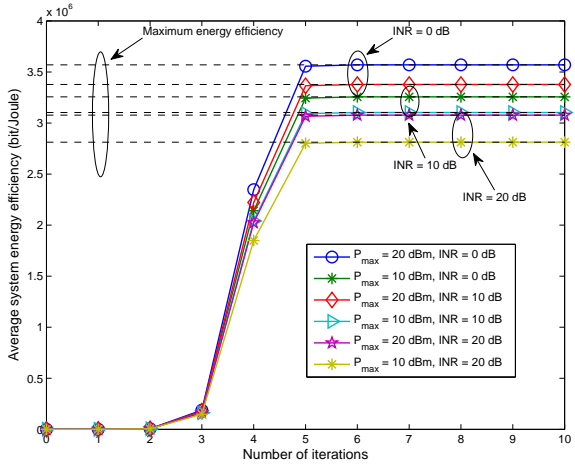}\vspace*{-4mm}
\caption{Average system energy efficiency (bit/joule) versus number of
iterations for different
 levels of INR, $\frac{\sigma_{I_{i}}^2}{\sigma_{z^s}^2}$, and different
values of maximum transmit power allowance, $P_{\max}$. The dashed
 lines
represent the maximum energy efficiency
for the different
cases. } \label{fig:convergence}\vspace*{-4mm}
\end{figure}

Figure \ref{fig:convergence} illustrates the evolution of the average system energy efficiency of  the
proposed iterative algorithm for different levels of average received interference. In particular, we focus on the convergence speed of the proposed algorithm for a given value of optimal $\rho$.  The results in Figure
\ref{fig:convergence} were averaged over 100000 independent
 realizations for   multipath fading. The dashed
lines denote the average maximum energy efficiency for each case study.
It can be observed that the iterative algorithm converges to the
optimal value within 5 iterations for all considered scenarios. Besides, the variations in the INR level $\frac{ \sigma_{I_{i}}^2}{\sigma_{z^s}^2}$  and the maximum transmit power allowance $P_{\max}$  have a negligible impact on the convergence speed of the proposed algorithm.

In the sequel, we set the number of iterations to 5 for illustrating the performance of the proposed algorithm.

\subsection{Average Energy Efficiency}
\begin{figure}[t]
 \centering\vspace*{-5mm}
\includegraphics[width=3.5 in]{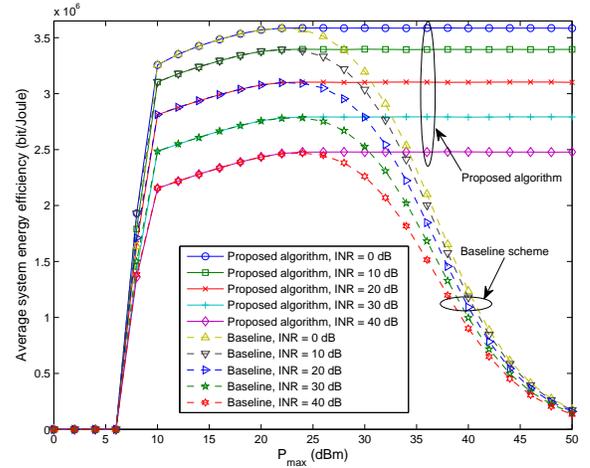}\vspace*{-4mm}
\caption{Average system energy efficiency (bit-per-Joule) versus
maximum transmit power allowance, $P_{\max}$, for different levels of INR, $\frac{\sigma_{I_{i}}^2}{\sigma_{z^s}^2}$.} \label{fig:EE_PT}\vspace*{-4mm}
\end{figure}

Figure \ref{fig:EE_PT} depicts the  average system energy efficiency versus the
maximum transmit power allowance, $P_{\max}$, for different received levels of interference.
 It can be
seen that for  $P_{\max}<10$ dBm, the system energy efficiency is zero since the optimization problem in (\ref{eqn:cross-layer}) is infeasible due to an insufficient power transmission in the RF for satisfying the constraints on $R_{\min}$ and $P_{\min}^{req}$.  However, for a large enough $P_{\max}$, the energy efficiency of the proposed algorithm first increases with increasing $P_{\max}$ and then approaches a constant as the energy efficiency gain due to a higher transmit power allowance gets saturated. This is because the transmitter is not willing to
consume an exceedingly large amount of power for RF transmission, when the maximum system  energy efficiency is achieved.
Furthermore, the energy efficiency of the system is impaired by an increasing amount of interference, despite the potential energy efficiency gain due to energy harvesting from interference signals, cf. (\ref{eqn:Power_harvested_d}) and (\ref{eqn:avg-sys-eff}). For comparison, Figure
\ref{fig:EE_PT} also contains the energy efficiency of a baseline
power allocation scheme in which the system
capacity (bit/s) with constraints C1--C6 in
(\ref{eqn:cross-layer}) is maximized. It can be observed
that for the low-to-moderate maximum transmit power allowance regimes,
i.e., $P_{\max}<24$ dBm, the baseline
scheme achieves the same performance  as the proposed algorithm in terms of energy efficiency.
This result indicates that in the low transmit power allowance
regime, an algorithm which achieves the maximum system capacity may also achieve the maximum energy efficiency and vice versa. However, the energy efficiency of the
baseline scheme decreases dramatically in the high transmit power allowance
regime. This is because the baseline scheme employs a large transmit power for capacity maximization which is detrimental for energy efficiency maximization.

\subsection{Average System Capacity }

\begin{figure}[t]\vspace*{-5mm}
\centering
\includegraphics[width=3.5 in]{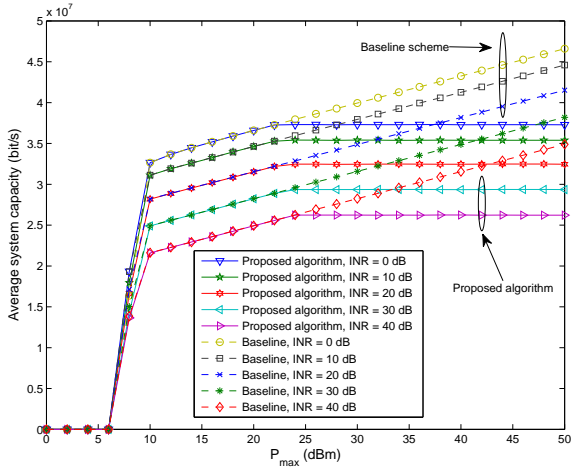}\vspace*{-4mm}
 \caption{Average system capacity (bit-per-second) versus
maximum transmit power allowance, $P_{\max}$, for different  levels of INR, $\frac{\sigma_{I_{i}}^2}{\sigma_{z^s}^2}$.} \label{fig:CAP_PT}\vspace*{-6mm}
\end{figure}

Figure \ref{fig:CAP_PT} shows the average system capacity versus
maximum transmit power allowance $P_{\max}$ for different levels of INR, $\frac{\sigma_{I_{i}}^2}{\sigma_{z^s}^2}$.  We compare the proposed algorithm again with the baseline scheme described in the last section.
The average system capacities of both algorithms are zero for  $P_{\max}<10$ dBm due to the infeasibility of the problem. For $P_{\max}>10$ dBm,  it
can be observed that the average system capacity of the proposed
algorithm approaches a constant in the high transmit power allowance regime.
This is because the proposed algorithm stops to consume more power
 for transmitting radio signals
  for maximizing the system energy efficiency. We note that, as expected, the baseline
scheme achieves a higher average system capacity than the proposed
algorithm in the high transmit power allowance regime. This is due to the fact that the baseline scheme consumes a larger amount of
transmit power compared to the proposed algorithm. However, the baseline scheme achieves the maximum
system capacity by  sacrificing the system energy efficiency.
\vspace*{-0.05cm}
\subsection{Average Harvested Power and Power Splitting Ratio}
\begin{figure}[t]\vspace*{-5mm}
 \centering
\includegraphics[width=3.5 in]{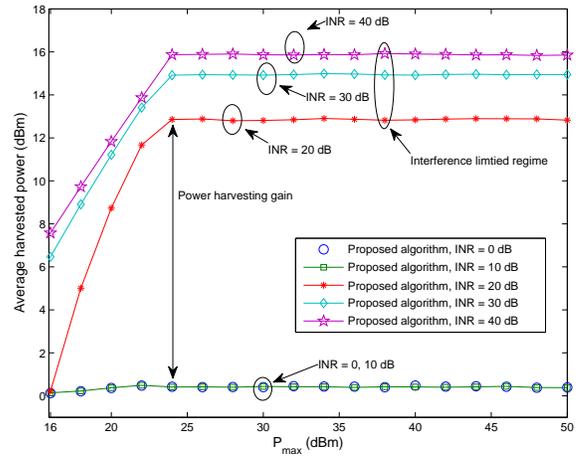}\vspace*{-4mm}
\caption{Average harvested power (dBm) versus
 maximum transmit power allowance, $P_{\max}$,  for different levels of INR, $\frac{\sigma_{I_{i}}^2}{\sigma_{z^s}^2}$. The double-sided arrow indicates the power harvesting gain due to an increasing $\rho$ in the interference limited regime, cf. Figure \ref{fig:rho}.  } \label{fig:power_harvested}\vspace*{-3mm}
\end{figure}

\begin{figure}[t]
 \centering\vspace*{-1mm}
\includegraphics[width=3.5 in]{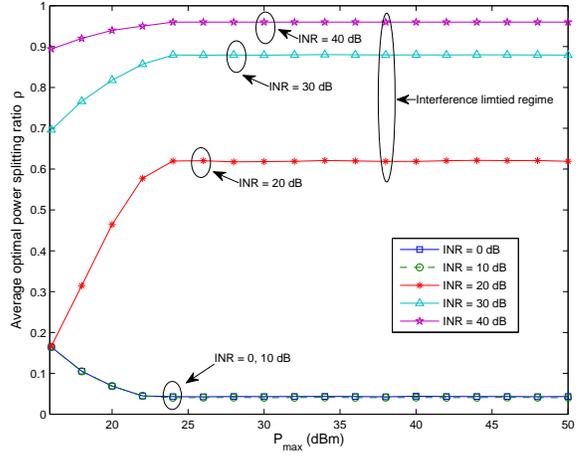}\vspace*{-4mm}
\caption{Average optimal power splitting ratio, $\rho$, versus
 maximum transmit power allowance, $P_{\max}$,  for different  levels of INR, $\frac{\sigma_{I_{i}}^2}{\sigma_{z^s}^2}$.} \label{fig:rho}\vspace*{-6mm}
\end{figure}

Figures \ref{fig:power_harvested} and \ref{fig:rho} show, respectively, the average harvested power and the average optimal power splitting ratio, $\rho$, of the proposed algorithm versus
maximum allowed transmit power, $P_{\max}$, for different levels of INR, $\frac{\sigma_{I_{i}}^2}{\sigma_{z^s}^2}$.
It can be observed in Figure \ref{fig:power_harvested} that for small values of INR, i.e., $\mbox{INR}\le10$ dB, only a small amount of power is harvested by the receiver for energy efficiency maximization. In other words, a small portion of received power is assigned to the energy harvesting unit, cf. Figure \ref{fig:rho}. In fact, for  small values of INR, assigning a larger amount of the received power for information decoding provides a higher  capacity gain to the system  which results in an improvement in energy efficiency. On the contrary, as  shown in Figure  \ref{fig:rho}, the receiver has a higher tendency to assigning a larger portion of  the received power to the energy harvester  in the interference limited regime, i.e., $\mbox{INR}\gg 10$ dB.  Indeed, the SINR on each subcarrier approaches a constant in the interference limited regime and is independent of $\rho$, i.e, $\frac{(1-\rho)l
g\abs{H_i}^2P_i}{(1-\rho)(\sigma_{z^a}^2+\sigma_{I_i}^2)+\sigma_{z^s}^2}\rightarrow \frac{P_i l
g\abs{H_i}^2}{\sigma_{I_i}^2+\sigma_{z^a}^2}$. Thus, assigning more received power for information decoding does not provide a significant  gain in channel capacity. On the contrary, the total power consumption decreases linearly w.r.t. an increasing $\rho$. As a result,   assigning a larger portion of the received power to energy harvesting can enhance the system energy efficiency when the capacity gain is saturated in the interference limited regime.

\vspace*{-0.1cm}
\section{Conclusions}\label{sect:conclusion}
In this paper, we formulated the power allocation
algorithm design for simultaneous  wireless information and power transfer in OFDM systems as a non-convex optimization problem. In the problem formulation, we took into account a minimum data rate
requirement, a minimum required power transfer, and  the circuit power
dissipation. The multi-dimensional optimization problem was solved by using non-linear fractional programming, dual decomposition, and a one-dimensional full search. The simulation results reveal an interesting trade-off
between energy efficiency, system capacity, and wireless power transfer.

\vspace*{-0.05cm}
\bibliographystyle{IEEEtran}
\bibliography{OFDMA-AF}

\end{document}